\documentstyle[preprint,aps]{revtex}
\begin{document}
\newcommand{\spg}{${\,\Sigma ^{+}\rightarrow p\,\gamma\,}$}
\newcommand{\spp}{${\,\Sigma ^{+}\rightarrow p\,\pi^{0}\,}$}
\newcommand{\snp}{${\,\Sigma ^{+}\rightarrow n\,\pi^{+}\,}$}
\newcommand{\spb}{${\,\overline{\Sigma} ^{-}\rightarrow
\overline{p}\,\pi^{0}\,}$}
\newcommand{\csg}{${\,\Xi ^{-}\rightarrow \Sigma ^{-}\,\gamma\,}$}
\newcommand{\clk}{${\,\Xi ^{-}\rightarrow \Lambda ^{0}\,K ^{-}\,}$}
\newcommand{\kpp}{${\,K ^{-}\rightarrow \pi^{-}\,\pi^{0}\,}$}
\newcommand{\ocg}{${\,\Omega ^{-}\rightarrow \Xi ^{-}\,\gamma\,}$}
\newcommand{\ocpm}{${\,\Omega ^{-}\rightarrow \Xi ^{0}\,\pi ^{-}\,}$}
\newcommand{\olk}{${\,\Omega ^{-}\rightarrow \Lambda ^{0}\,K ^{-}\,}$}
\newcommand{\ocp}{${\,\Omega ^{-}\rightarrow \Xi ^{-}\,\pi ^{0}\,}$}
\newcommand{\ocx}{${\,\Omega ^{-}\rightarrow \Xi ^{-}\,X\,}$}

\draft

\title{New Upper Limit for the Branching Ratio of the
$\Omega^{-} \rightarrow \Xi^{-} \gamma$ Radiative Decay.}
\author{I.~F.~Albuquerque$^{(1),(a)}$,
N.~F.~Bondar$^{(3)}$, R.~Carrigan Jr.$^{(2)}$, D. Chen$^{(9),(b)}$,
P.~S.~Cooper$^{(2)}$, Dai~Lisheng$^{(4)}$, A.~S.~Denisov$^{(3)}$,
A.~V.~Dobrovolsky$^{(3)}$, T.~Dubbs$^{(7,g)}$,
A.~M.~F.~Endler$^{(11)}$, C.~O.~Escobar$^{(1)}$, M.~Foucher$^{(13),(c)}$,
V.~L.~Golovtsov$^{(3)}$, H.~Gottschalk$^{(2),(i)}$,
P.~Gouffon$^{(1),(d)}$, V. T.~Grachev$^{(3)}$,
A.~V.~Khanzadeev$^{(3)}$, M.~A.~Kubantsev$^{(8)}$, N.~P.~Kuropatkin$^{(3)}$,
J.~Lach$^{(2)}$, Lang~Pengfei$^{(4)}$,
Li~Chengze$^{(4)}$, Li~Yunshan$^{(4)}$,
M.~Luksys$^{(10)}$, J.~R.~P.~Mahon$^{(1),(d,h)}$,
E.~McCliment$^{(7)}$, A.~Morelos$^{(2),(e)}$, C.~Newsom$^{(7)}$,
M.~C.~Pommot~Maia$^{(12),(f)}$, V.~M.~Samsonov$^{(3)}$,
V.~A.~Schegelsky$^{(3)}$,
Shi~Huanzhang$^{(4)}$, V.~J.~Smith$^{(5)}$, Tang~Fukun$^{(4)}$,
N.~K.~Terentyev$^{(3)}$, S.~Timm$^{(6)}$, I.~I.~Tkatch$^{(3)}$,
L.~N.~Uvarov$^{(3)}$, A.~A.~Vorobyov$^{(3)}$, Yan~Jie$^{(4)}$,
Zhao~Wenheng$^{(4)}$, Zheng~Shuchen$^{(4)}$, Zhong~Yuanyuan$^{(4)}$}
\author{(E761 Collaboration)}
\address{$^{(1)}$ Universidade de Sao Paulo, Sao Paulo, Brazil}
\address{$^{(2)}$ Fermi National Accelerator Laboratory, Batavia, IL 60510}
\address{$^{(3)}$ Petersburg Nuclear Physics Institute, Gatchina, Russia}
\address{$^{(4)}$ Institute of High Energy Physics, Beijing, PRC}
\address{$^{(5)}$ H.H. Wills Physics Laboratory, University of Bristol, UK}
\address{$^{(6)}$ Carnegie Mellon University, Pittsburgh, PA 15213}
\address{$^{(7)}$ University of Iowa, Iowa City, IA 52242}
\address{$^{(8)}$ Institute of Theoretical and Experimental Physics, Moscow,
Russia}
\address{$^{(9)}$ State University of New York at Albany, Albany, NY  12222}
\address{$^{(10)}$ Universidade Federal da Paraiba, Paraiba, Brazil}
\address{$^{(11)}$ Centro Brasileiro de Pesquisas Fisicas, Rio de Janeiro,
Brazil}
\address{$^{(12)}$ Conselho Nacional de Pesquisas CNPq, Rio de Janeiro,
Brazil}
\address{$^{(13)}$ J.W. Gibbs Laboratory, Yale University, New Haven, CT 06511}
\date{\today}
\maketitle
\begin{abstract}

We have searched for the rare hyperon radiative decay \ocg , using the
Fermilab Proton Center 375 GeV/c charged hyperon beam.
Our measurement of the $\Omega^{-}$ beam fraction at 13.6 m from production
is (4.8 $\pm$ 0.4)$\times 10^{-5}$. No signal was found and
we determine a new upper limit of $4.6 \times 10^{-4}$ at 90\% CL for the
\ocg\ branching ratio.
\end{abstract}

\narrowtext

Among the six hyperon radiative decays with $\Delta S = 1$, the \ocg\ presents
a special
challenge to both experimentalists and theorists. This is because
of the low flux of $\Omega^{-}$ when compared to other
hyperons and the theoretical expectation of having the smallest branching
ratio of all the hyperon radiative decays \cite{kogan}. To the theorist the
challenge is that this is the only such decay involving a transition from an
SU(3) decuplet to an octet with the inherent complications of a spin 3/2
system. It has been conjectured \cite{eeg,kamath} that this decay is a good
testing ground for penguin diagrams whose magnitude remains uncertain.

The unitarity lower bound for the branching ratio \cite{kogan} of the \ocg\ is
$0.8 \times 10^{-5}$. An estimate \cite{kogan} of the real part of the
amplitude leads to a branching ratio of (1.0-1.5) $\times 10^{-5}$.
The only previous experimental information \cite{bourquin} for this decay is
an upper limit of $2.2 \times 10^{-3}$.

The E761 experiment was performed at the Fermilab Proton Center beam line
during the
1990 fixed target run. The main goals of the experiment were to determine the
asymmetry parameter and the branching ratio for the weak hyperon radiative
decays \spg\ \cite{moe} and \csg\ \cite{tim}.
The 800 GeV/c proton beam from the Tevatron impinged on a 1 interaction
length Cu target to produce a 375 GeV/c charged hyperon beam. This target
was placed at the beginning of the channel inside the Hyperon Magnet
(Fig.~\ref{fig:layout}).

The spectrometer was divided into three main systems in order to measure the
trajectory and momentum of the hyperon, the baryon and the photon energy.
The hyperon spectrometer consisted of 3 stations of silicon strip
detectors (SSD) with 50 $\mu$m pitch and one dipole magnet with a field
integral of 4.3 T$\cdot$m. The resolutions ($\sigma$) obtained, for this part
of the
spectrometer, were $\Delta p$/p = 0.7 \% , 12 $\mu$rad
and 5 $\mu$rad for momentum, horizontal and vertical angles respectively.

The 12 m long decay region (Fig.~\ref{fig:layout}) starting at 14 m from the
target was filled with helium gas.
The baryon spectrometer consisted of 4 stations of multiwire proportional
chambers (MWPC) and 2 dipole magnets (a total of 5.3 T$\cdot$m). A third
magnet (2.6 T$\cdot$m) was situated downstream of the last MWPC in order to
deflect low momentum
baryons which were part of the background to another decay analysed by E761,
the  \csg\ \cite{tim}. The resolutions of the baryon spectrometer were
$\Delta$p/p = 0.6\%, 21 $\mu$rad and 12 $\mu$rad for momentum, horizontal and
vertical angles respectively.

The photon spectrometer had two steel plates \cite{foot1} and a photon
calorimeter which consisted of lead glass and bismuth germanate (BGO) blocks.
The steel plate converted the photon, whose energy was measured in the
calorimeter ($\Delta$E/E $\sim$ 30\%/$\sqrt{E(GeV)}$ plus a 3\% constant term
added
in quadrature). This system had a square hole in its center to allow the baryon
through. A more detailed description of the E761 apparatus can be found in
references \cite{moe,ivone} \cite{foot2}.

The trigger required a beam particle through a coincidence of scintillation
counters, a photon conversion in one of the steel plates
and a high momentum baryon. The latter was defined by scintillation counters.
Undecayed beam particles were rejected by counter V1 and low momentum baryons
by V2 (see fig.~\ref{fig:layout}). We collected 2.8 $\times 10^{8}$ events
with this trigger.

The data analysed here were selected after requiring good quality hyperon and
baryon track reconstruction, as well as a decay vertex in the decay region.
We measured the
missing neutral mass squared ($M_{x^{0}}^{2}$) under the hypothesis that the
hyperon was an $\Omega^{-}$ and the baryon a $\Xi^{-}$.

The backgrounds consisted of the decays \spb , \kpp\ and \ocp . These
occur at a much higher rate than the \ocg\ decay and have a
$\pi^{0}$ in the final state which could satisfy the trigger. The first two
can be separated using as kinematical variables the ratio between the baryon
and hyperon momentum and the angle between their trajectories. The other
source of background, \ocp , is harder to separate due to the small mass
difference of the $\pi^{0}$ and $\gamma$ compared to the momentum of the
$\Omega^{-}$.

In order to separate \ocg\ from \ocp , we use information from the
photon calorimeter. The reconstruction of the hyperon and baryon
momenta and tracks
allows us to determine the position where the neutral particle would hit
the photon calorimeter. The single gamma from the \ocg\ has its energy
concentrated around the extrapolated neutral trajectory while the two gammas
coming from the $\pi^{0}$ decay deposit their energy over a larger region.
The energy of the single $\gamma$ is measured by the sum of the
energy on the block which is hit by the extrapolated neutral trajectory plus
the energy in the neighboring blocks. Thus the separation between
these two kinds of events is performed with a cut on the ratio between
this energy and the total energy in the photon calorimeter. We require this
ratio to be greater than 0.95.

Care must be taken when one of the $\gamma$'s coming from the
$\pi^{0}$ decay goes through the hole of the photon calorimeter. To
eliminate these events, we require the minimum value for the ratio between
the total energy in the photon calorimeter and the missing momentum (momentum
of the hyperon minus the momentum of the baryon) to be 0.80.

A further source of background is beam particle interactions. In order to
decrease their
number, we required that the decay vertex occurred between 1 to 8 meters from
the start of the the decay region, where the material in the beam is minimized.
 We also found that the horizontal
projection of the angle between the hyperon and baryon trajectories ($T_{X_B}$)
is smaller for interactions than for decays. The requirement that
$T_{X_B} > 2.4 \times 10^{-3}$mrad lowered the background by almost a factor
of 2.

Figure~\ref{fig:ocg} plots $M^{2}_{x^{0}}$, under the hypothesis \ocx , with
all the data that survived the cuts described above. Note the clear signal for
the \ocp\ but no signal for the \ocg . The solid line corresponds to the fitted
exponential background plus a gaussian at the $\pi^{0}$ peak.

Two steps are then taken, one
to confirm that there is no signal for the \ocg\ and the other to obtain
an upper limit with 90\% CL. The first consists of adjusting a
gaussian centered to the left of the $\pi^{0}$ peak (see Fig.~\ref{fig:ocg}),
by an amount
equal to the $\pi^{0}$ mass squared and with the same width \cite{foot3}
as the
\ocp\ peak. The amplitude of such a fit is negative which
assures us that there is no signal around the gamma mass squared. The second
step, in order to
obtain the number of events that gives us 90\% CL, is to force the amplitude
of the gaussian to change in such a way that the $\chi^{2}$ of the fit
will increase by 1.64 units. The result of this procedure is a total of 9
events.

The normalization needed for extracting an upper limit for the branching ratio
is provided by the
$\Omega^{-}$ fraction in our secondary beam. This was determined by the
analysis of the decays \olk\ \cite{foot4} and \ocp . Both analyses agree and
the results are dominated by statistical errors. Their average
gives an $\Omega^{-}$ fraction of $(4.8 \pm 0.4) \times 10^{-5}$ in the E761
hyperon beam, at 13.6 meters from the target. The average transverse momentum
of production is 0.66 GeV/c while the average $X_{F}$ is 0.47.

The upper limit for the branching ratio is given by:

\begin{equation}
BR < \frac{N_{\gamma}}{\varepsilon_{\gamma} \times f_{\Omega^{-}} \times
N_{bp}}
\label{eq:br}
\end{equation}
where $N_{\gamma}$ is the number of \ocg\ at 90\% CL,
$\varepsilon_{\gamma}$ is the efficiency for this decay, $f_{\Omega^{-}}$ is
the fraction of $\Omega$'s in the hyperon beam and $N_{bp}$ is the number of
particles in the hyperon beam. $\varepsilon_{\gamma}$ includes geometrical
acceptance,
trigger and cuts efficiencies. It was determined by Monte Carlo studies which
used the same experimental apparatus and the same trigger and cuts (except
those related to the photon calorimeter) as the data analysed. The efficiency
of the cuts related to the photon calorimeter was determined through studies
using samples of single photons obtained from the \spg\ analysis. Table 1 lists
the values used in Equation~\ref{eq:br}.

We obtain as a final result the following upper limit at 90\% CL:

\begin{eqnarray}
BR(\Omega^{-}\rightarrow \Xi^{-}\,\gamma) < 4.6 \times 10^{-4}
\end{eqnarray}

This result is smaller than the previous upper limit \cite{bourquin} by a
factor of 4.8. It confirms the indication \cite{tim} that radiative decays
of hyperons which do not contain a valence u quark have low branching ratios.
We note that this new measurement is still more than an order of magnitude
above the unitarity lower bound \cite{kogan}.

We wish to thank the staffs of Fermilab and the Petersburg
Nuclear Physics Institute for their assistance. The loan of the photon
calorimeter lead glass from Rutgers University is gratefully acknowledged.
This work is supported in part by the U.S. Department of Energy under contracts
DE-AC02-76CH03000, DE-AC02-76ER03075 and grants DE-FG02-91ER404682,
DE-FG02-91ER404664, DE-FG02-91ER404631,
the Russian Academy of Sciences, and the UK Science and Engineering Research
Council.

\begin{figure}
\caption{Plan view of the E761 apparatus in the Fermilab Proton Center charged
hyperon beamline.}
\label{fig:layout}
\end{figure}

\begin{figure}
\caption{Distribution of $M_{x^{0}}^{2}$ for the hypothesis \ocx . The
arrow indicates the $\pi^{0}$ squared mass. No signal appears on the gamma
squared
mass.}
\label{fig:ocg}
\end{figure}

\begin{table}
\caption{Parameters used for branching ratio upper limit in Equation 1.}
\vspace{2 mm}
\centering
\label{tab:br}
\begin{tabular}{lcc}
$N_{\gamma}$ & 9 (90\% CL)\\
$\epsilon_{\gamma}$ & 0.78 $\pm$ 0.02 \%\\
$f_{\Omega^{-}}$ & (4.8 $\pm$ 0.3) $\times$ $10^{-5}$\\
$N_{bp}$ & $5.21 \times 10^{10}$\\
\hline
BR $<$ & $4.6 \times 10^{-4}$ (90\% CL)\\
\end{tabular}
\vspace{4 mm}
\end{table}


\begin{references}
\bibitem{foot1} In the analysis of other decays studied by E761
a transition radiation detector (TRD) was used and the steel plates were then
needed. The TRD information was not used in this analysis.
\bibitem{foot2} Differences between the geometry for the \spg\ decay \cite{moe}
and the one
for the negative hyperons \cite{tim}, are described in \cite{moe,tim}
\bibitem{foot3} Monte Carlo studies show that the widths of the \ocp\ and
the \ocg\ distributions are the same.
\bibitem{foot4} To study this decay, we removed the photon calorimeter and
modified the trigger, to require a single beam particle and a hit (from the
$K^{-}$ in the V1 counter (fig.~\ref {fig:layout}).
\end{references}
\end{document}